# Split-pulse X-ray photon correlation spectroscopy with seeded X-rays from X-ray laser to study atomic-level dynamics

Yuya Shinohara[1*], Taito Osaka[2], Ichiro Inoue[2], Takuya Iwashita[3], Wojciech Dmowski[4], Chae Woo Ryu[4], Yadu Sarathchandran[5], Takeshi Egami[1,4,5]

[1]*Materials Science and Technology Division, Oak Ridge National Laboratory, Oak Ridge, Tennessee 37831, USA*

[2]*RIKEN SPring-8 Center, Sayo, Hyogo 679-5198, Japan*

[3]*Department of Integrated Science and Technology, Oita University, Dannoharu, Oita 870-1192, Japan*

[4]*Department of Materials Science and Engineering, The University of Tennessee, Knoxville, Tennessee, 37996 USA*

[5]*Department of Physics and Astronomy, The University of Tennessee, Knoxville, Tennessee, 37996 USA*

Corresponding author: Yuya Shinohara (shinoharay@ornl.gov)

**With their brilliance and temporal structure, X-ray free-electron laser can unveil atomic-scale details of ultrafast phenomena. Recent progress in split-and-delay optics (SDO), which produces two X-ray pulses with time-delays, offers bright prospects for observing dynamics at the atomic-scale. However, their insufficient pulse energy has limited its application either to phenomena with longer correlation length or to measurement with a fixed delay-time. Here we show that the combination of the SDO and self-seeding of X-rays increases the pulse energy and makes it possible to observe the atomic-scale dynamics in a timescale of picoseconds. We show that the speckle contrast in scattering from water depends on the delay-time as expected. Our results demonstrate the capability of measurement using the SDO with seeded X-rays for resolving the dynamics in temporal and spatial scales that are not accessible by other techniques, opening opportunities for studying the atomic-level dynamics.**

Since the observation of X-ray speckles using coherent X-rays,[1] scientists have developed a speckle-based technique called X-ray Photon Correlation Spectroscopy (XPCS), where the temporal correlation of speckle patterns is used to extract dynamics of materials at specific correlation lengths.[2–5] Earlier XPCS studies using synchrotron X-rays primarily focused on determining dynamics on relatively longer length-scales (> 10 nm).[6–11] Only a few studies have recently been carried out at the atomic-scales although the timescale is limited to a relatively slow dynamics (> 10 s).[12–14] Extending the timescale of XPCS to picoseconds, which are relevant to studying the atomic-level dynamics in matter such as liquid, has been challenging because of the insufficient intensity of coherent X-rays and the limited frame-rate of detectors (typically < 1 kHz). Meanwhile, recent progress in high-energy-resolution inelastic X-ray and neutron scattering has extended their accessible timescale and its applications.[15,16] Nevertheless, their timescale and length-scale are not appropriate for probing the atomic-level dynamics in a wide timescale of picoseconds to nanoseconds due to the limitation in energy resolution and intensity. Accordingly, the atomic-level dynamics in liquids, particularly supercooled liquids, remains elusive.

The advent of X-ray free electron lasers (XFELs),[17,18] together with recent progress in split-and-delay optics (SDO),[19,20] has raised the expectation for bridging the aforementioned gap by extending the timescale of XPCS to picoseconds.[21] In this timescale, measuring the time-correlation of each speckle is not practical because of the intrinsic limitation on the temporal resolution of X-ray detectors as well as the repetition rate of the XFELs. This limitation is overcome by a split-and-delay approach,[21,22] which uses two X-ray pulses that are generated from a single XFEL pulse. The accessible timescale is determined by the ability to generate two pulses separated in time $\Delta t$. The contrast $\beta(Q, \Delta t)$ at momentum transfer $Q$ in the sum of speckle scattering patterns from a sequence of two separate X-ray pulses is analyzed to extract the information on dynamics by the Speckle Visibility Spectroscopy (SVS),[23,24] utilizing the relationship between $\beta(Q, \Delta t)$ and the intermediate scattering function, $f(Q, \Delta t)$.[20,22,25–27] Recently SVS using X-ray pulses with variable pulse duration has been demonstrated for the femtoseconds dynamics in supercooled water without using SDOs,[28] and accelerator-based double X-ray pulse generations[29,30] are also accomplished; however, their pulse-duration or time-delay is restricted either to a limited range of orders of femtosecond or to discrete steps of hundreds of picoseconds. Alternatively, in an SDO system, a single XFEL pulse is divided into two pulses by a beam splitter, and a delay-time between the pulses is controlled by their path length difference. Hard X-ray SDO systems have hitherto been developed,[19,31–34] and a few experimental results using the SDO have been reported.[20,35] However, their application was still limited to a longer length-scale (> 1 nm) and the study of atomic-level dynamics at sub-nm scale using X-ray SVS (XSVS) with an SDO has not been reported because of the insufficient pulse energy of X-rays.

Here we report the XSVS result utilizing the SDO combined with the self-seeded X-rays[36] at SPring-8 Angstrom Compact free-electron Laser (SACLA).[18] The self-seeding of X-rays[36–38] generates narrow-band X-rays and thus can provide higher X-ray intensity after passing through the SDOs. The increase in the X-ray pulse energy at the sample position enables the application of XSVS at high $Q$ (> 1 Å$^{-1}$), which is relevant to the study of atomic-level dynamics.

**Results**

**Statistics of self-seeded X-rays after the SDO.** The XSVS experiment was carried out using the setup shown in Fig. 1. A reflection self-seeding[36] at a photon energy of 10 keV was employed. Each X-ray pulse was then split into two sub-pulses using a wavefront-division SDO, making use of Si(220) crystals.[31] The delay-time $\Delta t$ between the sub-pulses was controlled by the path length difference between the variable-delay branch and the fixed-delay branch. We changed $\Delta t$ between 0 and 2 ps, although the accessible range of $\Delta t$ is much longer.[31] The pulse energy of the variable-delay branch ($I_{\text{delayed}}$) and the fixed-delay branch ($I_{\text{fixed}}$) were monitored by the beam intensity monitors, which were installed in the Split-Delay Optics system.[39] The exit beams were overlapped and focused to a size of 0.7 μm (H) and 0.9 μm (V) at the sample position with X-ray mirrors.[40]

Figure 2a shows the histogram of X-ray pulse energy at the sample position, $I_{\text{tot}} = I_{\text{delayed}} + I_{\text{fixed}}$. The dashed line is a fit to the experimental data using a gamma density distribution function:[41]

$$P(I_{\text{tot}}) = \left(\frac{M}{\langle I_{\text{tot}}\rangle}\right)^M \frac{\exp(-MI_{\text{tot}}/\langle I_{\text{tot}}\rangle)I_{\text{tot}}^{M-1}}{\Gamma(M)}, \quad (1)$$

where $\Gamma(M)$ is the gamma function, $\langle I_{\text{tot}}\rangle$ represents the average X-ray pulse energy over shots, and $M$ is the number of modes. Our analysis yields $\langle I_{\text{tot}}\rangle = 7.7$ µJ and $M = 2.24$, which is close to the values at the Linac Coherent Light Source (USA) using 8 keV X-rays ($M = 2.35$).[35] The energy bandwidth of single-shot self-seeded X-ray (~2 eV in FWHM) improved the throughput of the SDO system, compared to the case using only self-amplified spontaneous emission (SASE), where the averaged bandwidth is around 30 eV and $\langle I_{\text{tot}}\rangle = 0.38$ µJ.[31] This significant improvement facilitates the XSVS at high $Q$, where the probabilities of multiple photons per pixel events can be too low to conduct the XSVS. The split ratio, $R = I_{\text{fixed}}/(I_{\text{fixed}} + I_{\text{delayed}})$, was distributed shot-by-shot, and 13% of the total events satisfy the condition of the ratio of $0.475 < R < 0.525$ (Fig. 2b). This distribution originated from shot-to-shot variations in the profile and position of the incident X-ray beam.

In our experiment, a continuous stream of water with a flow rate of 0.7 mL/min was irradiated by the X-rays. The diameter of the water stream was 50 µm, much larger than the variation in the beam position at the sample during the experiment, as shown in Fig. 2c and 2d. The time-courses of the X-ray beam position clarify that the absolute position of the X-ray beam on the sample, $X$ and $Y$, drifted ~ 2 µm in 3 hours. However, the relative position between two sub-pulses, $\Delta X$ and $\Delta Y$, remained stable enough to assume that the two sub-pulses overlapped with each other on the sample. This high stability of the SDO system is a crucial factor for carrying out the XSVS experiment successfully. The scattering from the sample was recorded by three multi-port charge-coupled devices (MPCCDs)[42] located 1 m downstream of the sample to cover $Q$-ranges shown in Fig. 3. The speckle size is estimated to be around 0.18 mm, which is larger than the pixel size of the MPCCDs, 50 µm.

**X-ray scattering of water using the SDO and seeded X-rays.** Figure 3a shows an example of raw CCD images taken around $Q = 2$ Å$^{-1}$ with $I_{tot} = 3.3$ µJ. Because of the low scattering cross section of water, the signal produced by photons is sparsely distributed. Figure 3c shows a one-dimensional scattering intensity profile, which was averaged over 35,899 shots, the averaged pulse energy of which was 7.9 µJ. A droplet algorithm[43–46] was employed to convert raw data into the digitized X-ray photon images (see *Methods*). The number of photons at each pixel per shot was 0.64 times the estimated value that was calculated using the result of high-energy X-ray diffraction.[47] This discrepancy could originate from the limited alignment accuracy, ~ 20 µm, of the sample position relative to the X-ray beam, hence a smaller irradiated volume. The result shows that the observed X-ray intensities predominantly represent scattering from the water.

We now discuss the sample heating induced by the first sub-pulse. Because of the high pulse energy, the sample temperature may have risen before the second sub-pulse hit the sample. Using the sample thickness, X-ray pulse energy, the specific heat of water, and the transmission of water at 10 keV, the temperature raise can be estimated to be several hundred Kelvins depending on $I_{fixed}$. To evaluate the possible heating effect, we binned our data based on $I_{fixed}$ and $\Delta t$ and then analyzed the scattering intensity around $Q = 2$ Å$^{-1}$ (Fig. 4a). When the delay time is zero or small (< 0.2 ps), the intensity profiles remain the same regardless of $I_{fixed}$. However, with a longer delay time and higher $I_{fixed}$, the peak-shift to high $Q$ is observed as shown in Fig. 4b. Using the temperature-dependence of peak position,[47] we estimated the amount of temperature rise $\Delta T$, as shown in Fig. 4c. Note that the observed scattering intensity profile is the sum of the two scattering intensity profiles, the one from the fixed-delay branch at $t = 0$ and the other from the variable delayed branch at $t = \Delta t$. Thus, the actual temperature rise can be twice as large as $\Delta T$ shown in Fig. 4c. The result shows that there exists a heating effect only after $\Delta t = 0.5$ ps with $I_{fixed} < 3$ µJ, but is much smaller than the simple estimation. On the other hand, $\Delta T$ at 1.0 ps and 2.0 ps shows similar values regardless of $I_{fixed}$. X-rays first excite electrons, and the electronic energy is transferred to molecular motion (phonons). The $\Delta t$-dependence suggests that the heat dissipation through phonons comes into effect after 1 ps, which calls for further experiments and simulations.

**Speckle contrast analysis.** From the statistics of observed photons, the visibility of the X-ray speckle pattern at specific $Q$ was calculated. Following the previous studies,[20,27,28,35,43] we assume that the probability for observing $k$ photons at a single-pixel is described by a negative binomial distribution with the average number of photons per pixel, $\mu$, when the number of scattered photons is low:[41]

$$P(k|\mu, M_s) = \frac{\Gamma(k + M_s)}{\Gamma(M_s)k!} \left(\frac{M_s}{M_s + \mu}\right)^k \left(\frac{\mu}{M_s + \mu}\right)^{M_s}. \quad (2)$$

Here $M_s$ is the number of modes in observed scattering images and is related to $\beta$ such that $\beta = 1/M_s$.[20,22,43] After we binned our data based on $I_{fixed}$ as in the last section, we determined the value of $\beta$ by using the log-likelihood ratio statistic of this distribution as demonstrated in earlier studies[20] as shown in Fig. 5. At $Q = 2$ Å$^{-1}$ with $I_{fixed} < 3$ µJ, as the delay-time increased, the visibility of speckle reduces, reflecting the dynamics of the sample. At delay-times longer than $\Delta t = 1.0$ ps, the values of $\beta$ agree with that of the baseline value that was determined by the uncorrelated beams without spatial overlap. The decrease in the contrast is comparable to those calculated from the result of inelastic X-ray scattering (IXS) of water as shown in Fig. 5.[48,49] Here, the result of IXS was vertically shifted and scaled using the baseline value and the XSVS result at $\Delta t = 0$. Although the heating effect by the first sub-pulse may affect the dynamics at $\Delta t > 0.5$ ps even with $I_{\text{fixed}} < 3$ µJ, the agreement between the contrast decrease and the IXS result suggests that the datapoints up to, at least, 0.2 ps with $I_{\text{fixed}} < 3$ µJ are reliable. This is the first time that the dependence of the speckle contrast on the delay-time was measured using an SDO system. On the other hand, the dependence of $\beta(Q = 2 \text{ Å}^{-1}, \Delta t)$ on $I_{\text{fixed}}$ shows no clear decaying behavior for $I_{\text{fixed}} > 7$ µJ because of the heating effect discussed above. Besides, we could not find a clear decaying behavior at $Q = 3$ Å$^{-1}$ or higher even with $I_{\text{fixed}} < 3$ µJ (Supplementary Fig. S1), presumably due to the smaller number of photons and the additional decoherence at a higher $Q$, which will be addressed in a future study.

**Discussion**

The estimated values of $M_s$ depend on (1) the dynamics of the sample to be obtained, (2) the splitting ratio $R$, (3) the initial contrast, (4) the contrast reduction due to the experimental setting, and (5) the degree of decoherence between the two split pulses.[20,22] In this letter, we did not include the data with $|R - 0.5| > 0.025$ to reduce the uncertainty with respect to (2). We can reasonably assume that the sample thickness involved in the scattering was constant because the shot-by-shot positional fluctuations of X-rays were small (Fig. 2c). Then, the effects of (3) and (4) can be expressed in terms of $\beta$, which was estimated to be $\beta_0 = 0.23 \pm 0.02$ at $Q = 2$ Å$^{-1}$ by measuring the contrast with a single pulse using only the fixed-delay branch by blocking the variable-delay branch. This value is consistent with the baseline value, because the baseline should be $\beta_0/2$.[20] Meanwhile, the result for dual pulses was $\beta(Q, \Delta t = 0) = 0.155 \pm 0.023$, smaller than $\beta_0$. This contrast reduction can be reasonably explained by the degree of decoherence between the sub-pulses. The angular mismatch parallel to the detector plane between the sub-pulses,[50] 0.15 mrad, creates the positional difference between two speckles ~ 0.15 mm on the detector plane, thereby reducing the contrast from 0.23 to ~0.15 ~ $\beta(Q, \Delta t = 0)$. Note that the angular mismatch perpendicular to the detector plane can be ignored. The perpendicular component of the mismatch is written as $-k_i \eta \sin 2\theta \sin \varphi + k_i O(\eta^2)$, where $k_i$, $\eta$, $2\theta$, $\varphi$ are the magnitude of the wavevector, the angular mismatch, the scattering angle, and the azimuthal angle of the scattering, respectively. In our case, $\varphi = 0$ and thus the perpendicular component is $k_i O(\eta^2)$, hence it is negligibly small. Shot-to-shot incomplete geometrical overlap of the sub-pulses could reduce the contrast.[20] The relative positional fluctuations were inherently random as shown in the lower rows of Fig. 2c and cannot be monitored simultaneously during the X-ray SVS measurement. Including these effects in analyses to provide reasonable estimates of decoherence using more sophisticated approaches such as hierarchical models[51] warrant future studies. The use of hierarchical models will also allow us to include the data with different $R$ for the estimation, thereby significantly increasing the statistics. This will make it possible to extract meaningful information at higher $Q$ and could facilitate the evaluation of femtoseconds to nanoseconds dynamics at the atomic scale.

The current results indicate that with $I_{\text{fixed}} > 3$ µJ the XSVS results of water in this $Q$-range are not reliable because of the sample heating. This does not mean that the increase in the X-ray pulse energy by the self-seeding was unnecessary. The number of shots with $1 < I_{\text{fixed}} < 3$ µJ was ~ 65 % of the total number of shots. Thus, most of the data was within the range where the heating effect has little effect at a timescale shorter than 0.5 ps. On the other hand, with $I_{\text{fixed}}$

< 1 µJ meaningful estimations of $\beta$ were not obtained because almost all the photon-counting event is not a multi-photon event but a single-photon event. Without the seeded X-rays, $\langle I_{\text{tot}} \rangle$ = 0.38 µJ,[31] and it was impossible to carry out the XSVS at the atomic-scale. Our results demonstrate the capability of XSVS measurement using the SDO with seeded X-rays as well as the advantages and necessity of the self-seeded X-rays in the XSVS at high-$Q$.

**Methods**

**Self-seeded X-rays.** We used an 8-GeV electron beam with a charge of 130 pC and ~10 fs duration for reflection self-seeding[36] using a Si(220) microchannel-cut crystal monochromator at SPring-8 Angstrom Compact free-electron Laser (SACLA)[18]. Eight undulator segments were used to generate the SASE with average pulse energy of 80 µJ, and a channel-cut crystal monochromator was used to select a 10 keV radiation with a bandwidth of 0.6 eV (FWHM). The X-ray was used as a seed, which was amplified by the 13 downstream undulator segments.

**Split–and-delay optics.** A wavefront-division SDO using Si(220) crystals was used to split a single pulse into two sub-pulses with a delay time.[31] The SDO system was installed 70 m downstream from the last undulator segment. Shot-to-shot noninvasive diagnostics of pulse energies for both branches were made using inline diagnostic modules. After propagating through the SDO, the pulse width of X-ray was ~8 fs, which was similar to that of SASE (6-8 fs).

**Experimental Setting**. The experiment was carried out at BL3, SACLA (Japan). The exit beams from the SDO system were focused on a sample position with a focusing mirror system and hit a water jet. The diameter of the water jet was 50 µm, and the water temperature was 22 ˚C. A CCD was located 0.3 m downstream of the sample for monitoring the overall beam intensity and position in shot-by-shot. Each pulse energy of the sub-pulses was monitored by the intensity monitors[39] in the SDO system. The time-courses of the X-ray beam position at the sample position were separately measured by installing another CCD with a magnification system at the sample position. Scattering from the water was recorded using three MPCCDs with a pixel size of 50 μm. The distance between the sample and the detector was 1 m. In this experimental setting, the speckle size on the detector plane is estimated to be 0.18 mm. As suggested by a recent study, an SDO with wavefront division has intrinsic reduction of speckle contrast due to the difference in the speckle position on the detector plane.[50] In the current setting, the angular difference between two sub-pulses was expected to be 0.15 mrad, leading the differences in the speckle positions between two sub-pulses ~ 0.15 mm on the detector plane, which corresponds to 83 % of the speckle size. This angular mismatch would lead to a 74% decrease in speckle contrast. Because the contrast with a single pulse from the fixed-delay branch was 0.23 and that with uncorrelated beam was 0.12, the decreased contrast would be (0.23-0.12) × 0.26 + 0.12 = 0.15, which is consistent with the measured values at $\Delta t = 0$ ps.

**Data Reduction.** In a CCD image, the electron charge cloud produced by an X-ray photon spreads over several pixels over the detector. A single 10 keV X-ray photon produces ~590 analog-to-digital units (ADU) on average, but the signal on a pixel is widely distributed due to the charge sharing between neighboring pixels, as shown in the solid line in Fig. 3b. We convert the raw data into the digitized X-ray photon images by using a droplet algorithm.[43–45] The histogram of the signal in a droplet in units of ADU is shown as the dashed line in Fig. 3b. Then, we carried out the photon assignment following a procedure labeled as Greedy Guess.[46] As discussed in great detail in reference,[46] the photon assignment process involves systematic errors but shows a linear response to contrast changes. In this work, we measured the baseline of the contrast, $\beta_0/2$, by shifting the beam position such that the two sub-pulses do not overlap with each other. $\beta_0$ was separately estimated by blocking the variable-delay branch. Then, the measured contrast was compared with these values without further calibration.

**Data availability**

Data supporting the findings of this study are available within the article and the supplementary materials and additional data are available from the corresponding authors upon reasonable request.

**Code availability**

The codes supporting the findings of this study are available from the corresponding authors upon reasonable request.

**Figures**

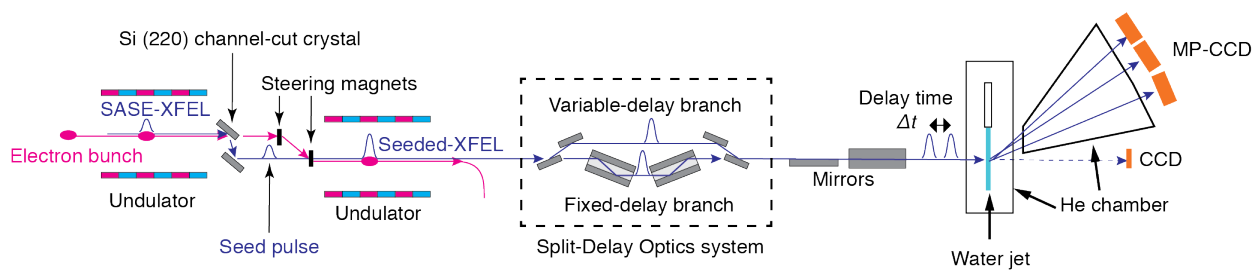

**Figure 1 | Experimental setting of XPCS using split-and-delay optics and reflection self-seeding at BL3, SACLA.**[18] The seed pulse was produced by monochromatizing the SASE from the upstream undulator segments. Then the seed was amplified in the downstream undulators. The details are described in ref.[36] The amplified X-ray pulse was split into two sub-pulses using

the SDO.[31] The sub-pulses with a delay time $\Delta t$ are focused by mirrors[40] and hit the water jet. Scattered X-rays were recorded by the MP-CCDs.[42]

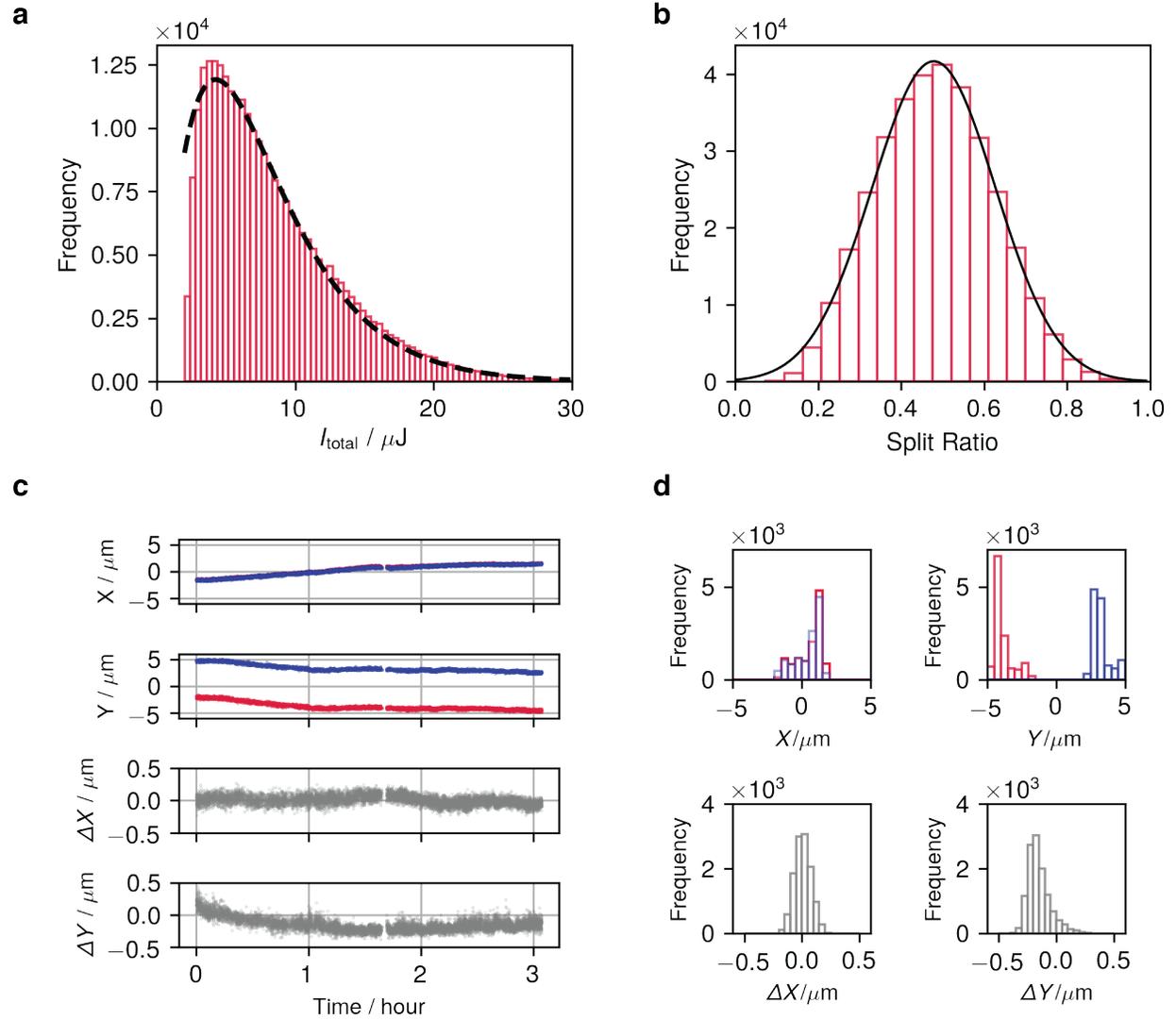

**Figure 2 | Statistics of self-seeded XFEL pulses after split–and–delay optics**. **a.** A pulse energy distribution at sample position for 280,815 shots (red bar). The dashed line represents the best fit to a gamma distribution. **b.** A histogram of the split ratio $R$ for 340,732 shots. The solid line represents the best fit to a Gaussian distribution for the average = 0.48 and the standard deviation = 0.15. **c.** Time-courses of X-ray beam position of two sub-pulses at the sample position, $X$ and $Y$, and their relative position, $\Delta X$ and $\Delta Y$. $\Delta X$ (or $\Delta Y$) were calculated by

subtracting $X$ (or $Y$) of the Variable-delay branch (blue) from $X$ (or $Y$) of the Fixed-delay branch (red). **d.** The histograms of $X$, $Y$, $\Delta X$ and $\Delta Y$. The position of two X-ray pulses was shifted in the $Y$ direction to distinguish them.

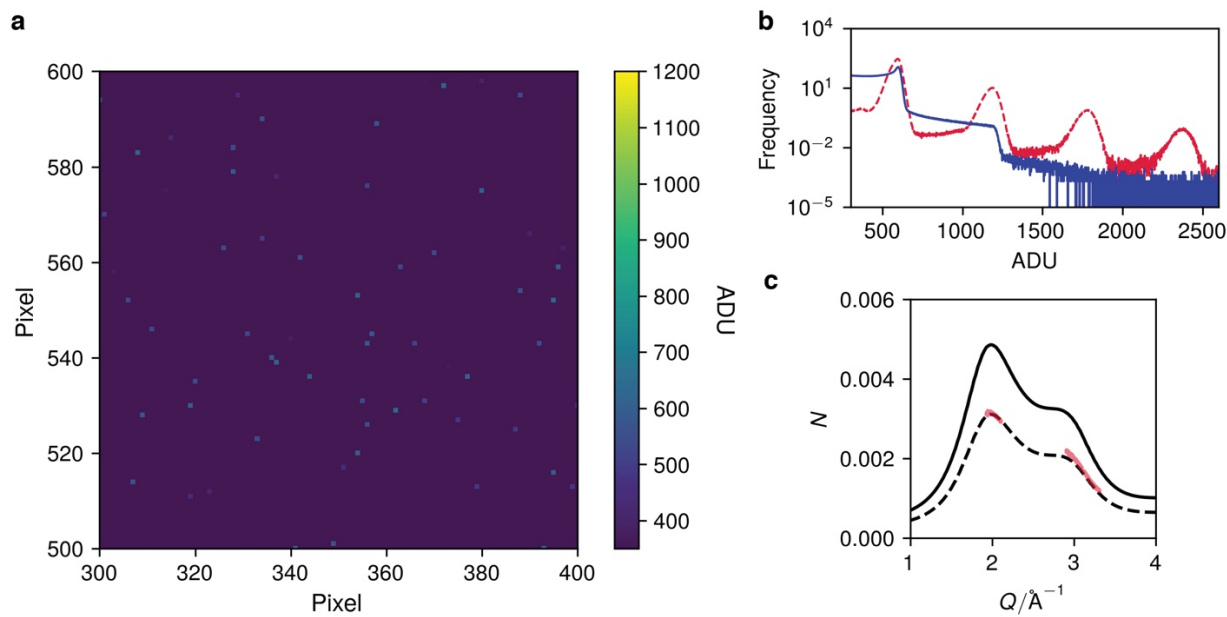

**Figure 3 | X-ray scattering from water**. **a.** Single split-pulse scattering pattern in a $100 \times 100$ pixel region of interest at around $Q = 2.0$ Å$^{-1}$ with $I_{tot} = 3.3$ µJ and $\Delta t = 0.0$ ps. The vertical and horizontal direction correspond to $Q$-direction and the azimuthal direction, respectively. **b.** Histogram of the charge (in units of analogue-to-digital units (ADU)) per shot in a $1024 \times 512$ pixel region. The blue solid line is the average histogram of the 287,982 raw images and the red dashed line is after the droplet algorithm has been applied. The single photon event corresponds to 595 ADU. **c.** The number of photons at each pixel per shot obtained by averaging 35,899 shots with $\Delta t = 0.0$ ps and $\langle I_{tot} \rangle = 7.9$ µJ. (red lines). The area around $Q = 2.0$ Å$^{-1}$ was covered by a single MPCCD and that around $Q = 3.0$ Å$^{-1}$ was covered by a dual MPCCD.[42] The solid black line is the estimation of X-ray elastic scattering using the result of high-energy X-ray diffraction

intensity profile,[47] the thickness of sample, the X-ray energy, the size of pixel, and the average number of X-ray photon for a single shot. The dashed line is × 0.64 of the solid line.

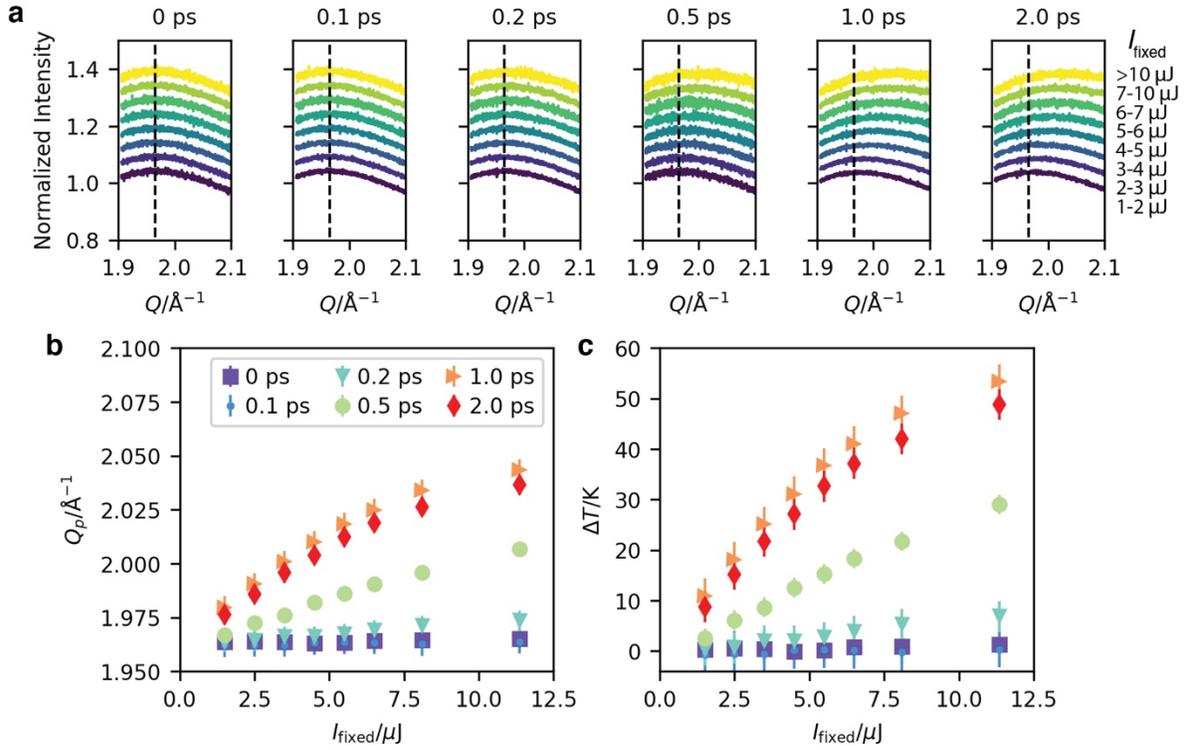

**Figure 4 | Dependence of scattering intensity profiles on the pulse energy of the fixed-delay branch**. **a.** Normalized scattering intensity around $Q = 2$ Å$^{-1}$. The delay time between two sub-pulses $\Delta t$ is shown at the top of the panels. The pulse energy of the fixed-delay branch $I_{\text{fixed}}$ is shown on the right. The profiles are vertically shifted for the sake of clarity. The dashed lines show the peak position when $\Delta t = 0$ and 1 μJ < $I_{\text{fixed}}$ < 2 μJ. **b.** Dependence of the peak position $Q_{\text{p}}$ on $I_{\text{fixed}}$. $\Delta t$ is shown as the legend. **c.** Estimated values of the temperature rise due to the sub-pulse from the fixed-delay branch. The error bars were determined by the standard deviation of the fitting of the peak position.

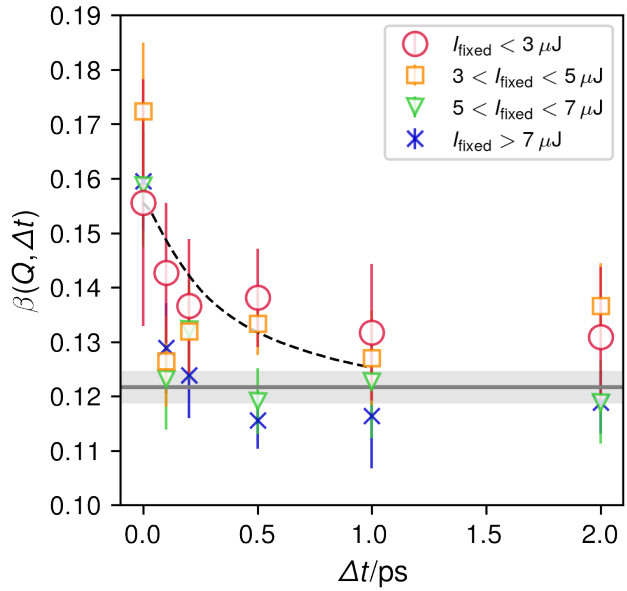

**Figure 5 | X-ray speckle contrast obtained by maximum-likelihood estimation**. At $Q = 2.00 \pm 0.06$ Å$^{-1}$ with (circles) $I_{fixed} < 3$ μJ, (squares) 3 μJ $< I_{fixed} < 5$ μJ, (triangles) 5 μJ $< I_{fixed} < 7$ μJ, and (crosses) $I_{fixed} > 7$ μJ. (Diamonds). The solid line represents the contrast measured when there was no overlap between two sub-pulses. The uncertainty was calculated using the second derivative of the log-likelihood based on the approach in reference[20] and the uncertainty for the solid line is represented by the shade. The dashed line represents the decaying behavior,[48,49] which are estimated by the result of inelastic X-ray scattering where their amplitude was set by the XSVS result at $\Delta t = 0$.

**Acknowledgments**

X-ray scattering work by Y. Shinohara, W.D., C.W.R., and T.E. was supported by U.S. Department of Energy, Office of Science, Office of Basic Energy Science, Division of Materials Sciences and Engineering. Work by T.O. was supported by JSPS KAKENHI (Grant No. 18K18307) and work by I.I. was supported by JSPS KAKENHI (Grant No. 19K20604). The experiments at SACLA were carried out under the approval of JASRI (Proposal No. 2018B8041, 2019A8043, and 2019B8011).

**Author Contributions**

Y. Shinohara, T.O., I.I., and T.E. conceived the project. T.O. and I.I. prepared the SDO and self-seeding of X-rays. Y. Shinohara, T.I., W.D., C.W.R. and Y. Sarathchandran carried out the experiments with support by T.O. and I.I. Y. Shinohara analyzed the experimental data. Y. Shinohara, T.O., I.I., and T.E. wrote the manuscript text. All authors reviewed and discussed the analysis and manuscript.

**Competing interests**

The authors declare no competing interests.

**Additional Information**

**Correspondence and requests for materials** should be addressed to Y. Shinohara (shinoharay@ornl.gov).